\documentclass[letterpaper, 10pt]{article}
\usepackage[utf8]{inputenc}
\usepackage[top=1in, bottom=1in, left=1in, right=1in]{geometry}
\usepackage{float}
\usepackage{color}
\usepackage{tikz}
\usepackage{booktabs}
\usepackage{comment}

\usepackage{titling}
\RequirePackage[explicit]{titlesec}
\titlespacing*{\section}{0pt}{0.5em}{0.3pt}
\titlespacing*{\subsection}{0pt}{0.35em}{0pt}
\titlespacing*{\subsubsection}{0pt}{0.25em}{0pt}

\usepackage{caption}
\RequirePackage{fancyhdr}
\usepackage{lastpage}
\usepackage{empheq}

\usepackage{mathpazo}
\usepackage{xcolor}
\definecolor{GT}{RGB}{0, 0, 0}
\definecolor{Conv}{RGB}{247,37,133}
\definecolor{MC}{RGB}{114,9,183}
\definecolor{MVN}{RGB}{67,97,238}
\definecolor{FE}{RGB}{76,201,240}

\usepackage[
  colorlinks=true,
  linkcolor=blue,
  anchorcolor=blue,
  citecolor=blue,
  filecolor=blue,
  menucolor=blue,
  runcolor=blue,
  urlcolor=blue]{hyperref}
\usepackage{amsmath, amsthm, amssymb, amsfonts}
\usepackage{mathrsfs}
\usepackage[version=4]{mhchem}
\usepackage{upgreek}
\usepackage{dsfont}
\usepackage{enumerate}
\usepackage{enumitem}
\usepackage{graphicx}
\usepackage{subcaption}
\usepackage{multirow}
\usepackage{multicol}

\usepackage[numbers, sort&compress]{natbib}

\makeatletter
\pretocmd\@maketitle
  {\def\@makefnmark{\hbox{\@textsuperscript{\normalfont\@thefnmark}}}}
  {}{\FAIL}
\makeatother
\makeatletter
\renewcommand*{\@fnsymbol}[1]{\ensuremath{\ifcase#1\or *\or \dagger\or \ddagger\or
   \mathsection\or \mathparagraph\or \|\or **\or \dagger\dagger
   \or \ddagger\ddagger \else\@ctrerr\fi}}
\makeatother

\title{Unsupervised Neural-Implicit Laser Absorption Tomography for Quantitative Imaging of Unsteady Flames}

\author{
  Joseph P. Molnar$^{1,}$\thanks{Authors made an equal contribution.}, Jiangnan Xia$^{2,3,*}$,
  Rui Zhang$^{2}$, Samuel J. Grauer$^{1}$, and Chang Liu$^{2,}$\thanks{Corresponding author: \href{mailto:c.liu@ed.ac.uk}{c.liu@ed.ac.uk}}\vspace*{.15em}\\
  {\small $^1$Department of Mechanical Engineering, Pennsylvania State University}\vspace*{-.25em}\\
  {\small $^2$School of Engineering, University of Edinburgh}\vspace*{-.25em}\\
  {\small $^3$School of Mechanical Engineering, Shanghai Jiao Tong University}}
\date{}

\begin{document}

\maketitle
\setcounter{footnote}{3}
\vspace*{-2em}

\begin{abstract}
\noindent 
This paper presents a novel neural-implicit approach to laser absorption tomography (LAT) with an experimental demonstration. A coordinate neural network is used to represent thermochemical state variables as continuous functions of space and time. Unlike most existing neural methods for LAT, which rely on prior simulations and supervised training, our approach is based solely on LAT measurements, utilizing a differentiable observation operator with line parameters provided in a standard spectroscopy database format. Although reconstructing scalar fields from multi-beam absorbance data is an inherently ill-posed, nonlinear inverse problem, our continuous space--time parameterization supports physics-inspired regularization strategies and enables data assimilation. Synthetic and experimental tests are conducted to validate the method, demonstrating robust performance and reproducibility. We show that our neural-implicit approach to LAT can capture the dominant spatial modes of an unsteady flame from very sparse measurement data, indicating its potential to reveal combustion instabilities in measurement domains with minimal optical access.\par\vspace{.5em}

\noindent\textbf{Keywords:} Laser absorption tomography,
quantitative imaging,
inverse problems,
neural-implicit reconstruction technique,
combustion diagnostics
\end{abstract}
\vspace*{2em}

\section{Introduction}
\label{sec:introduction}
Turbulent mixing and combustion are fundamental processes in power and propulsion systems \cite{Steinberg2021}. These applications often involve complex, unsteady flow and thermochemical fields that require advanced experimental techniques for accurate characterization. Spatio-temporally resolved data are essential for identifying turbulent structures. Optical diagnostics provide quantitative, non-intrusive one-, two-, and three-dimensional (1D, 2D, and 3D) measurements at high repetition rates, making them indispensable for such studies. However, many spatially resolved sensors, including those based on laser-induced fluorescence \cite{Stohr2012}, filtered Rayleigh scattering \cite{Schulz1996}, particle image \cite{Scarano2012} and tracking \cite{Schroder2023} velocimetry, or multi-angle light scattering \cite{Altenhoff2019}, require extensive optical access to the probe volume. This limitation confines their use to controlled laboratory environments. Additionally, these techniques often demand complex optics and precise calibration, making them vulnerable in harsh environments characterized by vibrations, high heating loads, or window fouling. Pilot- and full-scale power and propulsion applications thus necessitate a robust, high-speed imaging technique capable of operating under harsh conditions with minimal optical access. This need motivates the development of a neural-implicit algorithm for laser absorption tomography (LAT) \cite{Cai2017, Liu2019}, termed \emph{NILAT}: a robust approach for reconstructing challenging flows from sparse LAT data.\par

Laser absorption tomography employs multi-beam absorption spectroscopy to reconstruct 2D fields of mole fraction and temperature, and, in certain setups, pressure \cite{Grauer2019} or velocity \cite{Qu2018, Grauer2020}. LAT systems are flexible and accessible, often utilizing commercial laser diodes and requiring only a few pencil-sized entry and exit points for the laser beams. This minimal optical access has enabled the deployment of LAT across a wide range of power-generation and propulsion systems, including automotive \cite{Wright2010} and marine-engine pistons \cite{Tsekenis2017} for analyzing fuel-air mixing, industrial swirl combustors \cite{Liu2018} for assessing combustion efficiency and lean blowout limits, and gas turbine exhaust plumes \cite{Upadhyay2022} for monitoring carbon emissions. To achieve rapid imaging, LAT systems typically use a fixed array of beams. In small-scale setups, this may involve a few dozen beams, while larger systems have arrays with up to 150 beams \cite{Upadhyay2022, Jiang2023}.\par

Accurately inferring turbulent flow fields from LAT data has been a long-standing challenge \cite{Daun2016}. Typically, the region of interest (RoI) is represented using a pixel or triangle-element basis, with the measurement equations discretized accordingly. This approach results in a linear inverse problem for each wavenumber in spectrally resolved LAT or for each transition in spectrally integrated LAT. A set of linear reconstructions can then be locally post-processed to calculate thermochemical and velocity fields, as discussed in \ref{app:LAT:linear}. When the grid resolution is high enough to resolve turbulent structures, the system of equations becomes underdetermined because the number of basis functions, $n$, far exceeds the number of laser beams, $m \ll n$. This discrepancy necessitates regularization to produce unique, stable, and physically plausible reconstructions \cite{Daun2016}.\par

Iterative solvers, such as the algebraic reconstruction technique (ART) and its variants, have been widely used for decades \cite{Verhoeven1993}. These solvers exhibit semi-convergence, where the first few iterations capture robust low-frequency components of the solution, while later iterations are increasingly affected by noise \cite{Elfving2014}. Early stopping can yield reasonable but low-resolution estimates, effectively acting as a form of \emph{implicit} regularization. Optimizing the number of iterations is challenging, however, and ART reconstructions suffer from poor spatial fidelity compared to other algorithms. \emph{Explicit} regularization techniques are generally preferred due to their accuracy, predictable impact on solutions, and support for uncertainty quantification \cite{Kaipio2006}. In LAT, explicit methods typically impose spatial smoothness, either globally using Tikhonov regularization or locally using total variation regularization for edge preservation. Among explicit methods, Tikhonov regularization is particularly popular in the LAT community due to its simplicity, acceptable accuracy, and computational efficiency \cite{Dai2018, Wei2021a}. Comprehensive reviews of regularization techniques for LAT are provided by Cai \cite{Cai2017} and Liu \cite{Liu2019}.\par

Nonlinear methods for LAT directly parameterize reconstructions using temperature and mole fraction fields rather than absorbance fields, allowing regularization to be applied directly to these state variables \cite{McCann2022, Cai2017}. However, in its nonlinear formulation, LAT is an inherently non-convex problem, typically requiring a metaheuristic global optimization technique. This significantly increases the computational cost of tomographic reconstruction and, in some cases, degrades the solution quality. Most advancements in nonlinear LAT algorithms have focused on refining the optimization process rather than addressing the spatial characteristics promoted by the solver \cite{Ma2013, Dai2018, Shi2020}. Early nonlinear LAT algorithms employed spatial regularization strategies similar to those used in linear methods, such as Tikhonov or total variation regularization, as discussed in \ref{app:LAT:nonlinear}.\par

A new generation of nonlinear LAT algorithms leverages modern machine learning methods, generally categorized as supervised or unsupervised approaches. Supervised methods train a neural network using labeled input--output data pairs to directly map projection datasets to field variables like mole fraction and temperature \cite{Wei2021b, Jiang2023, Yu2018, Si2023}. These methods are typically trained on synthetic data, often generated using Gaussian phantoms or computational fluid dynamics (CFD) simulations. However, supervised methods face significant limitations, including the challenge of constructing representative training sets for complex, real-world combustion scenarios and the difficulty of generalizing to unseen flow and combustion features.\par

The second machine learning approach to nonlinear LAT employs an explicit measurement model to train a dedicated, target-specific neural network. Instead of learning a direct mapping from LAT data to field variables like temperature or mole fractions, these methods use a ``coordinate neural network'' to represent the gas as a function of spatial and temporal inputs, termed a neural-implicit representation. The network is trained by minimizing discrepancies between actual projection data and projections of the neural-implicit field variables, i.e., the hypothetical data corresponding to the current estimate encoded by the network. This approach, called the neural-implicit reconstruction technique (NIRT), has been successfully applied across various tomographic modalities, including X-ray radiography \cite{Ruckert2022}, emission imaging \cite{Kelly2024}, and schlieren-based techniques \cite{Molnar2024}. NIRT is versatile, accommodating any LAT measurements regardless of the sensor arrangement, and it does not rely on labeled training data.\par

Recently, Li et al. \cite{Li2024} proposed a NIRT algorithm for nonlinear LAT, representing the flow field with a simple coordinate neural network and using hyperspectral \ce{CO2} absorbance data to recover axisymmetric temperature and \ce{CO2} mole fraction fields in steady flames. Using a simple network (i.e., a standard multilayer perceptron, MLP) is effective for reconstructing smooth, steady fields because standard MLPs are biased toward low-frequency solutions: a form of implicit regularization. However, this NIRT approach struggles with low signal-to-noise ratios (SNRs) and sparse data for unsteady fields. A key challenge moving forward is to develop LAT methods capable of addressing more complex scenarios, such as reconstructing 2D distributions of temperature, mole fractions, and more in unsteady, turbulent flames with transient and asymmetric structures. Current linear and nonlinear LAT algorithms, including Li et al.'s NIRT technique, lack the neural expressivity needed to reconstruct high spatial frequency content from sparse measurements.\par

Our NILAT framework builds on the method of Li et al., incorporating an absorption spectroscopy measurement operator to compute synthetic projections from the network outputs. Unlike Li's algorithm, which performs radial reconstructions at discrete time instances, NILAT represents time-resolved field variables within the measurement plane, enabling a single-step ``$\text{2D} + t$'' reconstruction over an extended interval. This formulation allows the method to exploit spatio-temporal coherence in flow\slash combustion fields, improving reconstruction accuracy from sparse measurements. To represent complex, broadband dynamics, we augment the MLP with a Fourier encoding that enhances expressivity across spatial and temporal frequencies. However, with this added flexibility, explicit regularization becomes necessary to ensure physically plausible solutions: a consequence of the fundamentally ill-posed nature of LAT. We demonstrate that regularization is essential once the network is expressive enough to model unsteady fields, and we show that NILAT's optimization landscape is stable, enabling the use of classical techniques for optimal regularization such as L-curve analysis.\par

This paper outlines the fundamental principles of LAT, introduces our proposed reconstruction framework, and provides an analysis of regularization parameter selection. Section~\ref{sec:cases} outlines our selected flame configurations, while Sec.~\ref{sec:results} provides a parametric evaluation of NILAT, alongside experimental demonstrations involving small-scale burners, highlighting NILAT's ability to resolve unsteady flame dynamics.\par

\section{Neural-Implicit Laser Absorption Tomography}
\label{sec:LAT}
Laser absorption tomography extends laser absorption spectroscopy by utilizing multiple laser beams to capture spatially resolved information about a gas-phase species. By measuring light attenuation at various wavenumbers along these paths, LAT enables the inference of key properties such as temperature, chemical composition, velocity, and pressure. This section reviews the measurement model for absorption spectroscopy and introduces our neural reconstruction strategy for LAT. Lastly, we give an overview of regularization techniques.\par

\subsection{Absorption Spectroscopy Preliminaries}
\label{sec:NILAT:spectroscopy}
A fundamental quantity in absorption spectroscopy is the spectral absorbance,
\begin{equation}
    \alpha_\nu \equiv \log\mathopen{}\left(\frac{I_{0,\nu}}{I_\nu}\right) = \int_0^L \kappa_\nu\mathopen{}\left[ \mathbf{r}\mathopen{}\left(s\right) \right] \mathrm{d}s,
    \label{equ:BL}
\end{equation}
where $\nu$ is the detection wavenumber and $I_{0,\nu}$ and $I_\nu$ are the non-absorbing reference (flame-off) and attenuated (flame-on) intensities incident on the photodetector. The right-hand side of this expression follows from the Beer--Lambert law, where $\kappa_\nu$ is the local absorption coefficient, and the indicator function $\mathbf{r} : \mathds{R} \rightarrow \mathds{R}^2$ or $\mathds{R}^3$ represents the beam path. This function, $\mathbf{r}$, maps a progress variable, $s$, to a position along the beam of length $L$, defined such that $\lvert\mathrm{d}\mathbf{r}/\mathrm{d}s\rvert = 1$. Integrating over an absorption transition yields
\begin{subequations}
    \label{equ:int}
    \begin{align}
        \label{equ:int:absorbance}
        A_k &= \int_0^\infty \alpha_\nu \,\mathrm{d}\nu = \int_0^L K_k\mathopen{}\mathopen{}\left[ \mathbf{r}\mathopen{}\left(s\right) \right] \mathrm{d}s,
        \intertext{where}
        \label{equ:int:absorption}
        K_k &= \int_0^\infty \kappa_\nu \,\mathrm{d}\nu = S_k\mathopen{}\left(T\right) \frac{\chi p}{k_\mathrm{B} T}.
    \end{align}
\end{subequations}
Here, $A_k$ and $K_k$ denote the path-integrated absorbance and local absorption coefficient for the $k$th transition of the target species. The right-hand side of Eq.~\eqref{equ:int:absorption} connects the absorption coefficient to the thermodynamic state of the gas, where $S_k$ is the line strength for the $k$th transition, $T$ is the gas temperature, $\chi$ is the mole fraction of the target species, $p$ is the pressure, and $k_\mathrm{B}$ is the Boltzmann constant.\par

The line intensity for transitions of common gases in local thermodynamic equilibrium can be computed using line parameters from spectroscopy databases, such as HITRAN \cite{Gordon2022} or HITEMP \cite{Rothman2010},
\begin{equation}
    \label{equ:line intensity}
    S_k = S_{\mathrm{ref},k} \frac{Q\mathopen{}\left(T_\mathrm{ref}\right)}{Q\mathopen{}\left(T\right)}
    \frac{\exp\mathopen{}\left(-c_2 E_k^{\prime\prime}/T\right)}
    {\exp\mathopen{}\left(-c_2 E_k^{\prime\prime}/T_\mathrm{ref}\right)}
    \frac{1 - \exp\mathopen{}\left(-c_2 \nu_k/T\right)}
    {1 - \exp\mathopen{}\left(-c_2 \nu_k/T_\mathrm{ref}\right)}.
\end{equation}
In this expression, $T_\mathrm{ref}$ is a reference gas temperature (commonly 296~K), $S_{\mathrm{ref},k}$ is the line intensity at $T_\mathrm{ref}$, $Q$ is the total internal partition sums (TIPS) function, $c_2$ is the second radiation constant, $E_k^{\prime\prime}$ is the lower-state energy of the $k$th transition, and $\nu_k$ is the line center of the $k$th transition. The line center shifts based on the local thermochemical state and bulk gas velocity, a factor that must be accounted for in LAT when performing velocimetry. However, as this shift has a negligible effect on $S_k$, we approximate $\nu_k$ with the vacuum line center, $\nu_{0,k}$, in this paper.\par

\subsection{Neural-Implicit Reconstruction Technique}
\label{sec:NILAT:framework}
Equations~\eqref{equ:BL} to \eqref{equ:line intensity} enable calculation of the absorbance for a specific laser beam given: (i)~knowledge of the gas state, $(\chi, T, p)$, along the beam, (ii)~line parameters of the target molecule, $\{S_{\mathrm{ref},k}, E_k^{\prime\prime}, \nu_{0,k}\}$ for each measured transition, $k \in \mathcal{K}$, and (iii)~the molecule's TIPS function, $Q$. While line parameters and TIPS functions are readily available in databases such as HITRAN and HITEMP, the gas state is typically unknown and must be reconstructed from multi-beam absorbance data. This section introduces a neural-implicit reconstruction technique for LAT, referred to as NILAT.\par

We begin with a set of simultaneous absorbance measurements, denoted $A_{k,i}(t)$, where $k \in \mathcal{K}$ and $i \in \mathcal{I}$ indicate the absorption transition and laser beam index, respectively. These measurements are recorded at discrete time instances, $t_j$ for $j \in \mathcal{J}$. Our goal is to reconstruct continuous 2D distributions of the target mole fraction, $\chi$, and gas temperature, $T$, that match these measurements. To achieve this, we represent the gas using neural states,
\begin{equation}
    \label{equ:mapping}
    \mathsf{N} : \left(\mathbf{x}, t\right) \mapsto \left(\chi, T\right),
\end{equation}
where $\mathsf{N}$ is a deep feed-forward neural network that maps a spatial coordinate, $\mathbf{x}$, and time, $t$, to the quantities of interest. Details of the network architecture are provided in Sec.~\ref{sec:results:implementation} and \ref{app:network}; the framework can be extended to include additional state variables as needed.\par

The network is trained to reproduce measured data while conforming to prior information about the spatio-temporal dynamics of $(\chi, T)$. These objectives are encoded in a data fidelity term, $\mathscr{J}_\mathrm{data}$, a regularization penalty, $\mathscr{J}_\mathrm{reg}$, and an optional boundary penalty, $\mathscr{J}_\mathrm{bound}$. The total loss function is
\begin{equation}
    \label{equ:optimization}
    \mathscr{J}_\mathrm{total} = \mathscr{J}_\mathrm{data} + \mathscr{J}_\mathrm{reg} + \mathscr{J}_\mathrm{bound}.
\end{equation}
This aggregate loss is minimized via backpropagation, yielding a function $\mathsf{N}$ that fits the absorbance data while adhering to prior knowledge about $(\chi, T)$. For comparison, this study also evaluates a conventional LAT algorithm based on Tikhonov regularization, with details provided in \ref{app:LAT}.\par

The data loss is based on the absorption spectroscopy model outlined in Sec.~\ref{sec:NILAT:spectroscopy},
\begin{equation}
    \label{equ:data loss}
    \mathscr{J}_\mathrm{data} = \frac{1}{\left\lvert \mathcal{I} \times \mathcal{J} \times \mathcal{K} \right\rvert} \sum_{i \in \mathcal{I}} \sum_{j \in J} \sum_{k \in K} \left\{A_{k,i}\mathopen{}\left(t_j\right) -
    \int_0^{L_i} K_k\mathopen{}\left[\mathbf{r}_i\mathopen{}\left(s\right), t_j\right] \mathrm{d}s\right\}^2.
\end{equation}
Here, $A_{k,i}$ represents the absorbance of the $k$th transition measured by the $i$th laser at time $t_j$. The local absorption coefficient, $K_k$, is computed using Eqs.~\eqref{equ:int:absorption} and \eqref{equ:line intensity}, based on the $\chi$ and $T$ values predicted by $\mathsf{N}$ at the position $\mathbf{r}_i(s)$ and time $t_j$. The indicator function, $\mathbf{r}_i$, describes the path of the $i$th beam with a length $L_i$; the integral in Eq.~\eqref{equ:data loss} is approximated at each training iteration using Monte Carlo sampling.\par

To implement this model for a given molecule, the line parameters for selected transitions of the target species are specified, and the necessary quantities are evaluated using the expressions above. The TIPS function is computed via linear interpolation of tabulated values provided in increments of 1~K \cite{Gordon2022}. To ensure numerical stability during backpropagation in single precision and to reduce floating-point operations, the constant terms in Eqs.~\eqref{equ:int:absorption} and \eqref{equ:line intensity} are grouped together and their products are precomputed.\par

\subsection{Regularization Penalties}
\label{sec:NILAT:regularization}
The network $\mathsf{N}$ must possess sufficient expressivity to accurately represent the measured flow fields. While turbulent flows exhibit broadband spatial and temporal frequency content, gradient-descent-type training of a coordinate neural network inherently introduces a low-frequency spectral bias. To mitigate this, we include a Fourier encoding (detailed in \ref{app:network}) that enhances the network's ability to represent functions with broadband spectral content. However, a Fourier encoding can introduce random variations in $\chi$ and $T$ for limited-data tomography setups. Omitting the encoding, or using a smaller network, act as forms of implicit regularization: eliminating spurious high-frequency content but also inherently limiting the network's ability to represent the true fields. Since implicit regularization often has unpredictable effects on the solution, it is preferable to retain $\mathsf{N}$'s full capacity for capturing complex turbulent dynamics and instead incorporate explicit regularization, which imposes well-defined constraints with predictable outcomes to improve reconstruction accuracy, like spatial smoothness or known correlation length scales. The independent and combined effects of Fourier encodings and explicit regularization on reconstruction accuracy are documented in \ref{app:Fourier}.\par

In this work, we use a second-order Tikhonov penalty to produce smooth fields,
\begin{equation}
    \label{equ:L2 penalty}
    \mathscr{J}_g = \frac{1}{\left\lvert \mathcal{A} \times \mathcal{T} \right\rvert}\int_\mathcal{T} \int_\mathcal{A} \left\lVert\nabla^2 g\right\rVert_2^2 \mathrm{d}\mathbf{x} \,\mathrm{d}t,
\end{equation}
where $\mathcal{T}$ represents the measurement interval, $\mathcal{A}$ is the 2D or 3D RoI, $\nabla^2$ is the spatial Laplacian operator, $\left\lVert \cdot \right\rVert_2$ denotes the Euclidean norm, and $g$ refers to an output of $\mathsf{N}$ (either $\chi$ or $T$ in this case). Exact derivatives of the continuous field, $g$, are efficiently computed using automatic differentiation, and the integrals in Eq.~\eqref{equ:L2 penalty} are approximated via Monte Carlo sampling. The regularization loss is defined as
\begin{equation}
    \label{equ:regularization loss}
    \mathscr{J}_\mathrm{reg} = \gamma_\upchi \mathscr{J}_{\chi} + \gamma_\mathrm{T} \mathscr{J}_{T},
\end{equation}
where $\gamma_\upchi$ and $\gamma_\mathrm{T}$ are weighting parameters. Selection of these parameters is discussed in the next section.\par

A boundary loss can be incorporated to account for known ambient conditions at the periphery of the measurement domain, ensuring that the solution aligns with these conditions,
\begin{equation}
    \label{equ:boundary}
    \mathscr{J}_\mathrm{bound} = \frac{1}{\left\lvert \partial\mathcal{A} \times \mathcal{T} \right\rvert} \int_\mathcal{T} \int_{\partial\mathcal{A}} \gamma_\mathrm{bound,T} \left(T - T_0\right)^2 + \gamma_\mathrm{bound, \upchi} \left(\chi - \chi_0\right)^2 \mathrm{d}\mathbf{x} \ \mathrm{d}t.
\end{equation}
In this expression, $\partial\mathcal{A}$ denotes the boundary of the RoI and $T_0$ and $\chi_0$ are the known or estimated free-stream temperature and mole fraction, respectively. As in previous loss terms, these integrals are approximated using Monte Carlo sampling of the fields.\par

\section{Parameter Selection}
\label{sec:parameter selection}
Proper selection of the regularization parameters in Eqs.~\eqref{equ:regularization loss} and \eqref{equ:boundary} is crucial for accurate reconstruction \cite{Hansen1998}. For simplicity, consider the single-parameter case,
\begin{equation}
    \label{equ:2C loss}
    \mathscr{J}_\mathrm{total} = \mathscr{J}_\mathrm{data} + \gamma \mathscr{J}_\mathrm{reg}.
\end{equation}
Here, $\gamma$ governs the trade-off between minimizing measurement residuals and promoting the physics-inspired properties encoded in $\mathscr{J}_\mathrm{reg}$. Small values of $\gamma$ lead to non-physical, least-squared solutions, while large values overweight the regularization term, often resulting in overly smooth or uniform fields, as observed with Tikhonov regularization. The optimal value balances these competing objectives, producing an estimate that closely approximates the true (unknown) field. In practice, Eq.~\eqref{equ:2C loss} serves as reasonable surrogate for Eq.~\eqref{equ:optimization} in NILAT reconstructions of \ce{H2O}. This is because the mole fraction and temperature of water vapor are usually strongly correlated, and the normalization of their respective loss terms, $\mathscr{J}_\chi$ and $\mathscr{J}_T$, results in similar optimal weights, such that $\gamma = \gamma_\upchi = \gamma_\mathrm{T}$. Moreover, the boundary conditions are generally easy to satisfy, so $\gamma_\mathrm{bound,\upchi}$ and $\gamma_\mathrm{bound,T}$ require limited tuning.\par

\subsection{Classical Methods}
\label{sec:parameter selection:classical}
Numerous techniques have been developed to optimize $\gamma$. Phantom studies involve generating synthetic data from CFD simulations of a representative flow or flame using the experimental beam layout. The data are corrupted with noise and reconstructed across various $\gamma$ values; the optimal value is the one whose reconstructions best match the simulated ground truth. While effective, this approach is often impractical due to the difficulty of accurately simulating a relevant phantom. Another method, the discrepancy principle, posits that the data loss, $\mathscr{J}_\mathrm{data}$, should be of the same order of magnitude as the measurement noise variance \cite{Morozov1966}. However, this method often over regularizes solutions, resulting in smeared distributions. Generalized cross-validation (GCV) selects the largest value of $\gamma$ beyond which there is an inflection in the data loss, reflecting a trade-off between $\mathscr{J}_\mathrm{data}$ and $\mathscr{J}_\mathrm{reg}$ \cite{Golub1979}. While widely used, GCV is numerically sensitive, making it challenging to reliably identify the optimal parameter \cite{Hansen1998}.\par

The L-curve method provides a robust alternative. This approach involves plotting $\mathscr{J}_\mathrm{reg}$ against $\mathscr{J}_\mathrm{data}$ on logarithmic axes, forming an ``L'' shape. At small values of $\gamma$, $\mathscr{J}_\mathrm{data}$ is minimized while $\mathscr{J}_\mathrm{reg}$ remains large, and the opposite is true at large $\gamma$. The ``optimal'' value corresponds to the point of maximum curvature, representing the best compromise between the two losses \cite{Hansen1998}. This point can be visually identified or computed through finite differences or a singular value analysis. Similar to the L-curve, Daun proposed a singular value approach for discrete linear problems in which $\gamma$ is increased until the $m$th singular value of the aggregate operator, where $m$ is the number of beams, begins to rise \cite{Daun2010}. Loosely speaking, this criterion confines the effect of regularization to the null space of the measurement operator. While the L-curve, GCV, and Daun's method are conceptually related, the L-curve remains the most widely used parameter selection technique for LAT and is demonstrated for NILAT in Sec.~\ref{sec:results}.\par

\subsection{Auto-Weighting Methods}
\label{sec:parameter selection:auto}
A key challenge in regularization is the inconsistency between the regularization term, $\mathscr{J}_\mathrm{reg}$, and the physics of the target process. True fields do not minimize $\mathscr{J}_\mathrm{reg}$, except in trivial cases like uniform fields. As a result, regularization requires balancing two imperfect components: a data loss term, based on noisy measurements and an approximate operator, and a regularization term that does not fully align with the real system behavior. Adaptive weighting techniques, such as gradient-based and neural-tangent-kernel methods \cite{Wang2023}, have been proposed for physics-informed neural networks (PINNs). These methods aim to ensure that all loss terms contribute equally to parameter updates. In gradient-based auto-weighting, an objective loss of the form
\begin{equation}
    \label{equ:generic loss}
    \mathscr{J}_\mathrm{total} = \sum_i \gamma_i \mathscr{J}_i
\end{equation}
is periodically updated as follows:
\begin{equation}
    \gamma_i^\prime = \frac{\sum_j \left\lVert \nabla_{\boldsymbol\uptheta} \mathscr{J}_j \right\rVert_2}{\left\lVert \nabla_{\boldsymbol\uptheta} \mathscr{J}_i \right\rVert_2},
\end{equation}
where $\nabla_{\boldsymbol\uptheta}$ is the gradient operator with respect to the model parameters, $\boldsymbol\uptheta$. Hence, training is accelerated for slowly-decreasing loss components and damped for rapidly-decreasing components. Smoothing is applied to the update,
\begin{equation}
    \gamma_i^{k+1} = \beta \gamma_i^{k} + \left(1 - \beta \right) \gamma_i^\prime,
\end{equation}
where $\beta$ is the smoothing parameter and $\gamma_i^{k}$ is the weight for the $i$th loss term after $k$ iterations of training. This approach introduces three hyperparameters: the update frequency, smoothing factor, and initial $\gamma_i$ values.\par

In tomographic applications like LAT, $\mathscr{J}_\mathrm{data}$ and $\mathscr{J}_\mathrm{reg}$ are fundamentally inconsistent. This is unlike PINNs, where evolution of the true fields is assumed to align with the equations included in the physics loss. The inconsistency of loss components in tomography raises questions about the effectiveness of auto-weighting for LAT (and particularly for NILAT). We compare the classical L-curve method with adaptive weighting in Sec.~\ref{sec:results:phantom} to address this issue.\par

\section{Case Studies}
\label{sec:cases}
Four scenarios are analyzed in this work: one synthetic case and three experimental ones. The synthetic case is designed to mimic an unsteady flow of combustion products with spectral content similar to the experimental flows, all serving as benchmarks for evaluating NILAT. For the experimental cases, combustion products from three laboratory-scale burners are measured using a LAT sensor. Results from a conventional reconstruction algorithm are included in all tests to highlight the improved fidelity of NILAT.\par

\subsection{Laser Beam Array}
\label{sec:cases:array}
\begin{figure}[ht]
    \centering
    \includegraphics[height=2in]{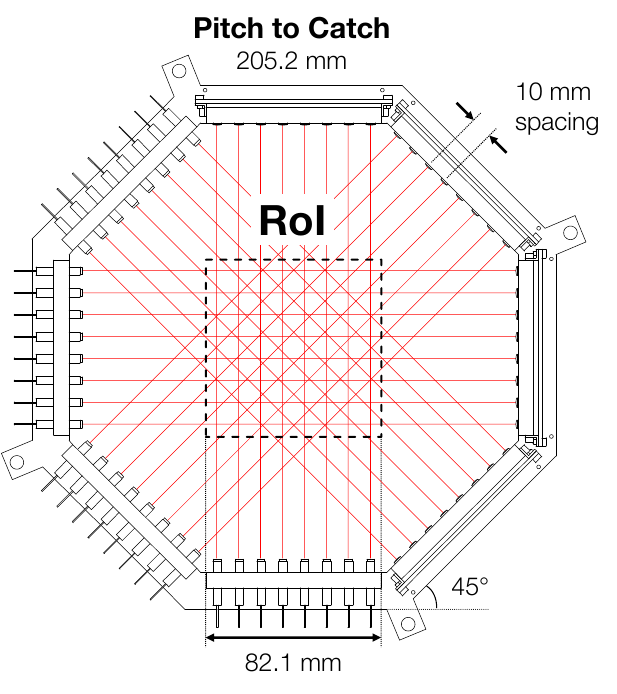}
    \caption{Schematic of the 32-beam LAT sensor used in the phantom study and experimental measurements.}
    \label{fig:Synth Setup}
\end{figure}

Figure~\ref{fig:Synth Setup} contains a schematic of the 32-beam LAT sensor and reconstruction domain used for our synthetic and experimental scenarios, alike. The RoI is a square area at the center of the sensor with an 82.1~mm edge length, corresponding to the LAT beams' common interrogation region. The sensor employs four banks of eight parallel beams, spaced 10~mm apart, with a 45$^\circ$ offset between banks \cite{Zhou2024}. The emitter and receiver units are separated by 205.2~mm. Light from two distributed feedback lasers is combined to probe \ce{H2O} transitions at 7185.59~cm$^{-1}$ and 7444.36~cm$^{-1}$ along each beam; these transitions were chosen for their differential sensitivity across the anticipated temperature range.\par

\subsection{Synthetic Dataset}
\label{sec:cases:phantom}
To mimic realistic turbulent flow features, we designed an analytical phantom with spatio-temporal variations representative of our experimental cases. Temperature and mole fraction fields are generated using circular Zernike polynomials \cite{Niu2022b}, fitted to distributions that mimic the combustion products of small-scale industrial and commercial burners. The mean fields have a toroidal structure, with peak values at the outer ring and lower values in the core, resembling a recirculation zone. Coherent temporal fluctuations are concentrated at the ``flame front,'' while incoherent spatio-temporal variations are distributed across the RoI. Z40 coefficients represent the mean and coherent components, with temporal oscillations modeled by a 9~Hz triangle-ramp function \cite{Niu2022b}. This introduces a spectral peak at 9~Hz and maximum coherent fluctuations of 150~K in temperature and $2.9 \times 10^{-3}$ in mole fraction. Pseudo-turbulence is generated with an additive Gaussian perturbation having a standard deviation of 2\% of the largest Z40 coefficient. This methodology produces seemingly turbulent behavior with prominent ``tonal'' fluctuations at the outer edge and a primary mode peaking at the prescribed frequency.\par

Note that both the phantom and NILAT estimates are continuous functions of $\mathbf{x}$, while the conventional algorithm represents the RoI on a $40 \times 40$-pixel grid. All fields are presented on that grid to make consistent quantitative comparisons. Phantoms are modeled as isobaric at 1~atm, with ambient conditions set to $T = 306$~K and $\chi_{\ce{H2O}} = 7.5 \times 10^{-3}$. The ambient region outside the RoI is uniform, and variations in $T$ and $\chi$ are perfectly correlated.\par

From the $T$, $p$, and $\chi$ fields, high-fidelity absorbance signals are generated using line parameters and TIPS functions from HITRAN2020 \cite{Gordon2022}. Absorbances are computed using Eq.~\eqref{equ:int:absorbance}, with the spatial integral approximated by sampling points along each line of sight between the emitter and receiver units. Pink additive noise with a standard deviation of 1\% of $\mathrm{max}(A_k)$ is added to the projection data to simulate realistic LAT imaging conditions. This noise level corresponds to an SNR of 40~dB, which is representative of laboratory or well-controlled industrial environments. Synthetic data are recorded at 250~Hz over a 10~s interval, and the resulting measurements align qualitatively with those observed in experimental flames, as shown in Sec.~\ref{sec:results:experiment}.\par

\subsection{Experimental Datasets}
\label{sec:cases:experiment}
\begin{figure}[ht]
    \centering
    \includegraphics[width=6in]{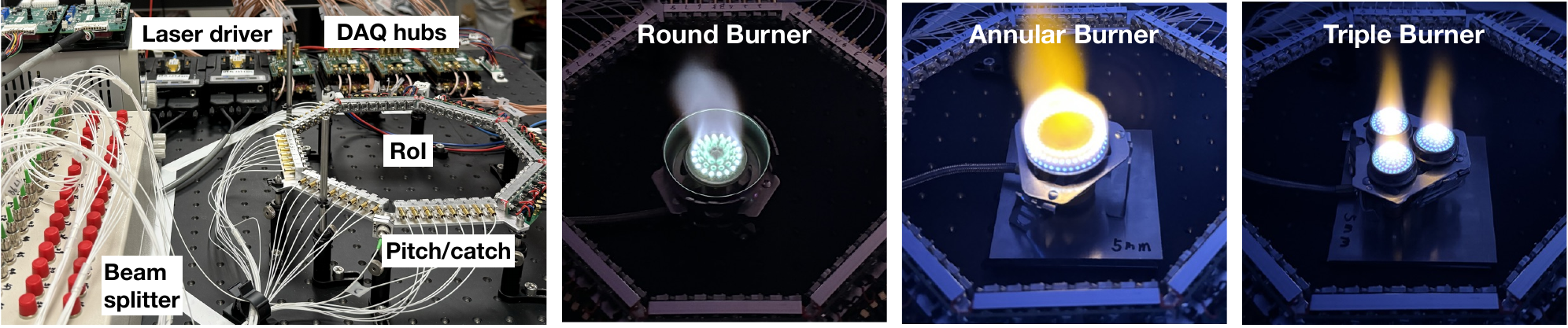}
    \caption{LAT sensor and data acquisition system (left panel), used to probe three commercially available burner configurations (right panel).}
    \label{fig:Expt Setup}
\end{figure}

Our experimental demonstrations are performed using the laboratory-scale burners shown in Fig.~\ref{fig:Expt Setup}. From left to right: (1)~the ``round burner'' has a 4.1~cm cap with closely spaced outlets across most of its surface, except for a small central region, producing a single round plume of combustion products; (2)~the ``annular burner'' features a 5.1~cm cap with outlets distributed across a sloping surface, having inner and outer diameters of 3.1 and 5.1~cm, generating a ring of flames; and (3)~the ``triple burner'' comprises three 2.6~cm caps with evenly spaced outlets, arranged in a triangular formation with a center-to-center spacing of 3.2~cm, producing three hot spots above the caps. The burners are fueled by propane, regulated via needle valves and pressure regulators, and measured using a mass flowmeter (Aalborg GFM17) at flow rates of 1.485, 1.099, and 1.103~L/min for the round, annular, and triple burners, respectively. Some air is entrained upstream of the outlets, resulting in a combination of partially- and non-premixed combustion. We estimate Reynolds numbers based on the individual nozzle diameters and the pure propane flow rate, yielding $Re_\mathrm{D} = 81.6, 34.3, \mathrm{and} \ 19.7$ for the round, annular, and triple burners, respectively. These values suggest laminar, relatively stable flames.\par

Measurements are taken at planes located 3~mm above the annular and triple burners and 7~mm above the round burner. The measurement planes were positioned close to the burner caps to capture relatively stable fields. The round burner includes an integrated wind shield that obstructs the optical path at 3~mm above the cap, necessitating a 7~mm measurement plane for this case. Each laser diode (NTT Electronics NLK1E5GAAA, NLK1B5GAAA) is temperature- and current-controlled by a laser driver (Wavelength Electronics LDTC 2-2E). Wavelength modulation is performed at a 1~kHz scan rate, with the 7185.59~cm$^{-1}$ and 7444.36~cm$^{-1}$ lasers multiplexed in the frequency domain using sinusoidal modulations at 100~kHz and 130~kHz, respectively. Each laser beam is collected by a photodetector and digitized with 16-bit resolution at 15.625~MS/s. All channels are synchronized by an external trigger and 4-to-1 multiplexed across neighboring scans, yielding an imaging rate of 250~Hz \cite{Zhou2024}, which is identical to the imaging rate in our phantom study. Path-integrated absorbances, $A_k$, as defined in Eq.~\eqref{equ:int:absorbance}, are calculated for each scan and beam by spectral fitting of the $2f/1f$ signal \cite{Goldenstein2014}. Measurements span a 10~s interval, during which the flames burn continuously at fixed propane flow rates. Following the LAT measurements, an S-type thermocouple is used to probe average temperatures at selected points above each burner. While these thermocouple measurements are intrusive and and biased by radiative heating of the probe, they provide a useful baseline to gauge the accuracy of our reconstructions.\par

\section{Tomographic Reconstructions and Analysis}
\label{sec:results}

\subsection{Implementation}
\label{sec:results:implementation}
Neural reconstructions were implemented in PyTorch using the architecture described in \ref{app:network}. Conventional algebraic reconstructions were computed as a baseline for comparison. For the two-step linear algorithm detailed in \ref{app:LAT}, the RoI was discretized into a $40 \times 40$-pixel grid, with uniform conditions applied outside the RoI. These ambient parameters, $(T_0, \chi_0)$, were optimized during the reconstruction. Local absorption coefficient fields, $K_k$, for the 7185~cm$^{-1}$ and 7444~cm$^{-1}$ transitions, were reconstructed using second-order Tikhonov regularization. Optimal regularization parameters were determined through an L-curve analysis. Reconstructed $(K_1, K_2)$ values were converted to $(T, \chi)$ at each pixel through ratiometric thermometry \cite{Goldenstein2013}.\par

The reconstructed fields were analyzed using spectral proper orthogonal decomposition (SPOD) \cite{Towne2018}. SPOD decomposes time-varying data into orthogonal modes ranked by energy, providing eigenvalues and spatial eigenvectors at selected frequencies to capture coherent spatio-temporal content in the dataset. In this work, SPOD was applied to the time-resolved temperature field estimates. Each analysis included all 2500 snapshots, which were recorded at 250~Hz. We used blocks of 250 time instances with 50\% overlap for SPOD, resulting in 19 blocks and a frequency resolution of 1~Hz.\par

\subsection{Phantom Study Results}
\label{sec:results:phantom}
We begin by analyzing results from the phantom study, focusing on the use of an L-curve to estimate the optimal regularization parameter. For simplicity, we use a single regularization weight, $\gamma = \gamma_\mathrm{T} = \gamma_\upchi$, enabled by proper normalization of the loss terms $\mathscr{J}_T$ and $\mathscr{J}_\chi$ using representative variances. The boundary losses, $\gamma_\mathrm{bound,T}$ and $\gamma_\mathrm{bound,\upchi}$, are readily satisfied when their weights exceed a minimal threshold.\footnote{Our results were invariant to the selection of $\gamma_\mathrm{bound,T}$ and $\gamma_\mathrm{bound,\upchi}$ across four orders of magnitude.} and thus require no tuning. Consequently, selecting an appropriate value for $\gamma$ in Eq.~\eqref{equ:2C loss} becomes the primary task for regularization in NILAT. All errors reported in this section are normalized root-mean-square errors.\par

\subsubsection{Parameter Selection Methods}
\label{sec:results:phantom:selection}
\begin{figure}[ht]
    \centering
    \includegraphics[width=0.9\textwidth]{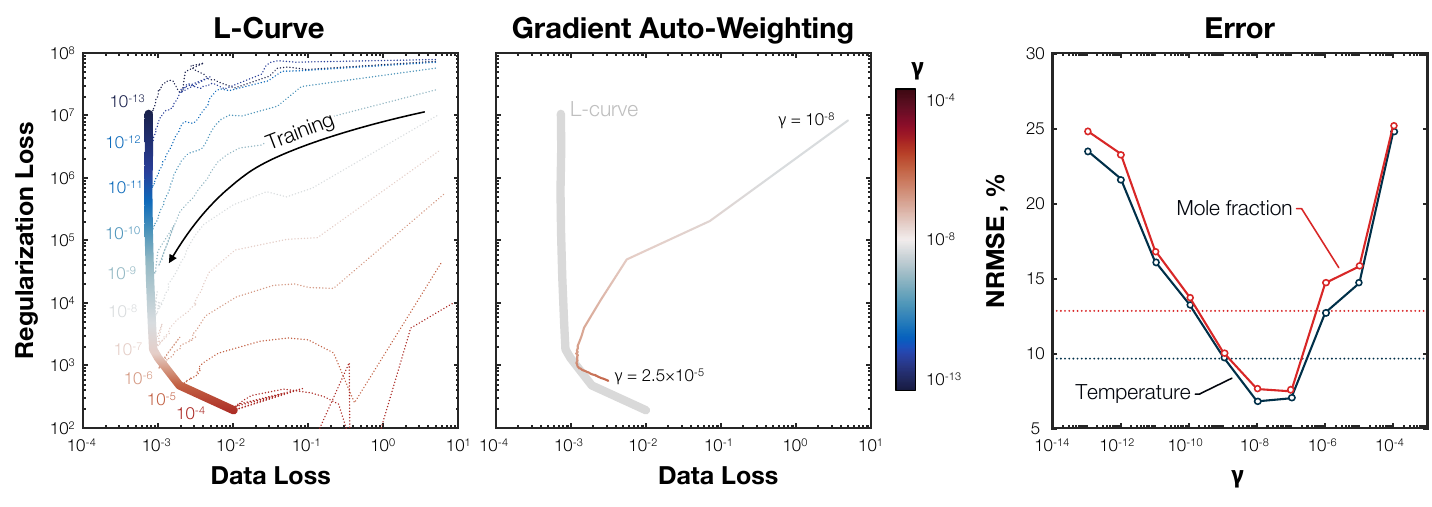}
    \caption{L-curve and auto-weighting behavior. Reconstructions at various values of $\gamma$ form an L-curve (left panel). The gradient-based auto-weighting trajectory initially approaches the optimal balance but diverges with continued training (center panel). Minimum reconstruction error aligns with the point of maximum curvature (right panel); dashed lines indicate errors from the conventional method.}
    \label{fig:L-curve}
\end{figure}

To explore the effects of regularization, we reconstructed the phantom using nine decades of $\gamma$ values ranging from $10^{-13}$ to $10^{-4}$. The leftmost plot in Fig.~\ref{fig:L-curve} illustrates the training progression for each case. At the outset, the randomly-initialized networks yield large values of $\mathscr{J}_\mathrm{data}$ and $\mathscr{J}_\mathrm{reg}$, placing them towards the upper-right corner of the plot. During training, the networks progress leftward and downward, as both losses are minimized, and converge to their respective terminus on the L-curve.\par

\begin{figure}[ht]
    \centering
    \includegraphics[width=0.9\textwidth]{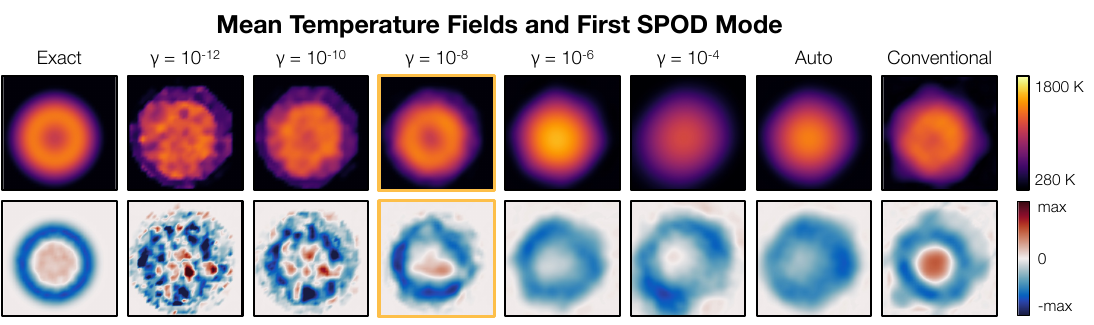}
    \caption{Reconstructions and SPOD modes for varying $\gamma$. (Top row) Sample reconstructions corresponding to different $\gamma$ values, the auto-weighted trajectory, and the conventional method. (Bottom row) First SPOD modes computed from each reconstructed dataset.}
    \label{fig:samples}
\end{figure}

Sample reconstructions of $T$ and the associated SPOD modes for several values of $\gamma$ are shown in Fig.~\ref{fig:samples}, while the rightmost plot in Fig.~\ref{fig:L-curve} illustrates reconstruction errors for $T$ and $\chi$ as functions of $\gamma$. Both fields exhibit similar behavior and error trends, supporting the use of a single regularization parameter since there is no trade-off between their accuracies. Reconstruction errors are minimized near $\gamma = 10^{-8}$ and $\gamma = 10^{-7}$, and the L-curve curvature is maximized near the latter point. Finer spacing of $\gamma$ values, particularly near the optimal region, would improve the precision of corner identification. Moreover, as noted in previous studies \cite{Hansen1998}, the L-curve method can sometimes lead to over-regularization, so the approach should be used with caution and supplemented with a phantom study. In this case, however, it serves as a good guide for selecting $\gamma$, especially since the low-error basin spans $\gamma = 10^{-8}$ to $10^{-7}$, offering some flexibility in parameter selection. These values can be interpreted as the optimal magnitude for balancing the contributions of noisy projection data with smoothness-based regularization. For scenarios where the reconstructed fields exhibit distinct behaviors, such as differing energy spectra, it may be necessary to optimize multiple independent regularization parameters, e.g., $\gamma_\mathrm{T}$ and $\gamma_\upchi$.\par

In addition to L-curve analysis, we evaluated the gradient auto-weighting technique proposed by Wang et al. \cite{Wang2023} for PINNs, as detailed in Sec.~\ref{sec:parameter selection:auto}. The central plot in Fig.~\ref{fig:L-curve} compares an auto-weighted trajectory with the L-curve, using an update frequency of 500 iterations, a smoothing factor of $\beta = 0.9$, and an initial $\gamma$ shown to be optimal in the rightmost panel of Fig.~\ref{fig:L-curve} ($\gamma = 10^{-8}$). While the auto-weighted trajectory initially approached the L-curve's point of maximum curvature, it diverged with continued training, bending toward the high-$\gamma$ leg. This behavior was observed across all tested hyperparameter settings and produced overly-smooth reconstructions, as shown in the ``Auto'' column of Fig.~\ref{fig:samples}. Divergence occurs because the magnitude of gradients of the regularization term, $\nabla_{\boldsymbol\uptheta} \mathscr{J}_\mathrm{reg}$, diminishes more rapidly than for the data fidelity term, $\nabla_{\boldsymbol\uptheta} \mathscr{J}_\mathrm{data}$. Smooth fields are inherently easier for the network to generate than fields consistent with the absorbance measurements, which drives repeated increases in $\gamma$. These increases progressively prioritize the regularization penalty, causing training to focus on minimizing $\mathscr{J}_\mathrm{reg}$, further reducing the magnitude of $\nabla_{\boldsymbol\uptheta} \mathscr{J}_\mathrm{reg}$. This creates a feedback loop that amplifies the emphasis on regularization while neglecting the data fidelity term, ultimately leading to suboptimal reconstructions. While this issue is less problematic in settings with consistent loss terms, it poses a challenge in tomographic applications where the loss components are inherently inconsistent. Gradient-based auto-weighting tends to disproportionately minimize one (inconsistent) loss component over the others, leading to imbalanced solutions. These findings suggest that traditional methods like the L-curve are preferable for selecting $\gamma$ in tomographic applications.\par

\begin{figure}[ht]
    \centering
    \includegraphics[width=5.5in]{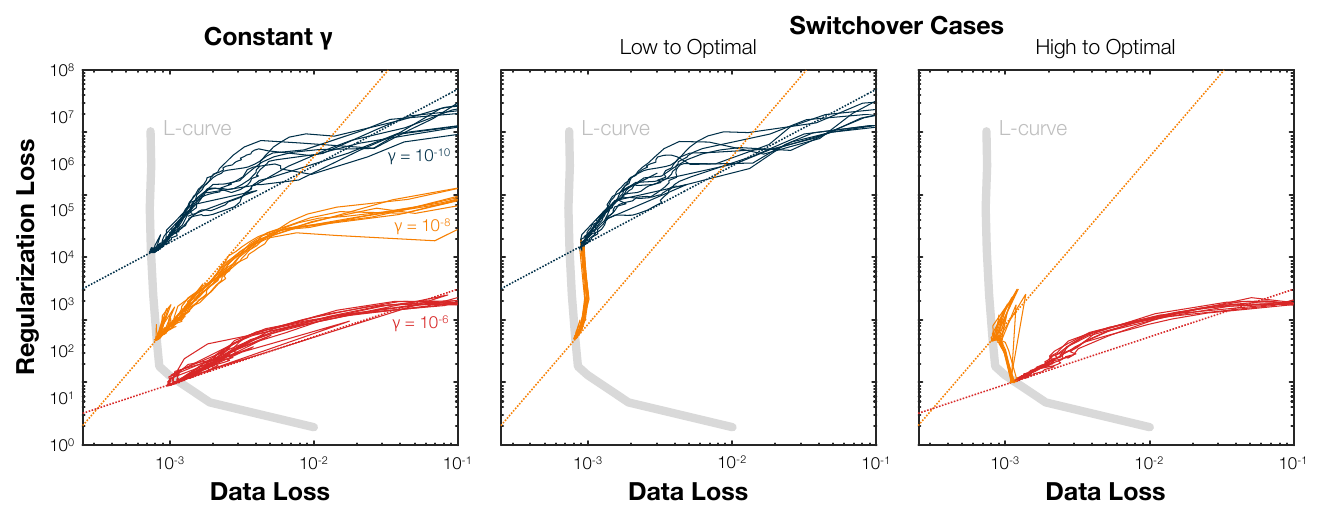}
    \caption{Training behavior under fixed and switching $\gamma$. Ensemble training trajectories show consistent convergence for fixed $\gamma$ values (left panel). When $\gamma$ is switched mid-training, networks quickly adapt to the new trajectory (center and right panels).}
    \label{fig:hysteresis}
\end{figure}

\subsubsection{Hysteresis Effects}
\label{sec:results:phantom:hysteresis}
Another important finding is that NILAT does not exhibit hysteresis effects during training, meaning the process is not path-dependent and converges to a stable optimum. This property is consistent with theoretical findings on loss landscapes for deep neural networks \cite{Choromanska2015}, and it simplifies the application of classical parameter selection methods. To test this, we trained networks with and without switching $\gamma$ halfway through optimization. Results are shown in Fig.~\ref{fig:hysteresis}. For each condition, an ensemble of ten networks was trained using one of three initial $\gamma$ values: $10^{-10}$, $10^{-8}$ (near the optimal value), or $10^{-6}$. In the left plot of Fig.~\ref{fig:hysteresis}, $\gamma$ was held constant during training, while in the middle and right plots, $\gamma$ was switched midway from $10^{-10}$ to $10^{-8}$ and from $10^{-6}$ to $10^{-8}$, respectively. In the latter cases, the networks quickly adjusted to the new $\gamma$ trajectory, following the corresponding path towards the terminus for $\gamma = 10^{-8}$. This behavior demonstrates that NILAT responds predictably to changes in $\gamma$, further supporting its compatibility with L-curve analysis and other classical parameter selection techniques.\par

\subsubsection{Assessing Reconstructions}
\label{sec:results:phantom:reconstructions}
\begin{figure}[htp]
    \centering
    \includegraphics[width=3.5in]{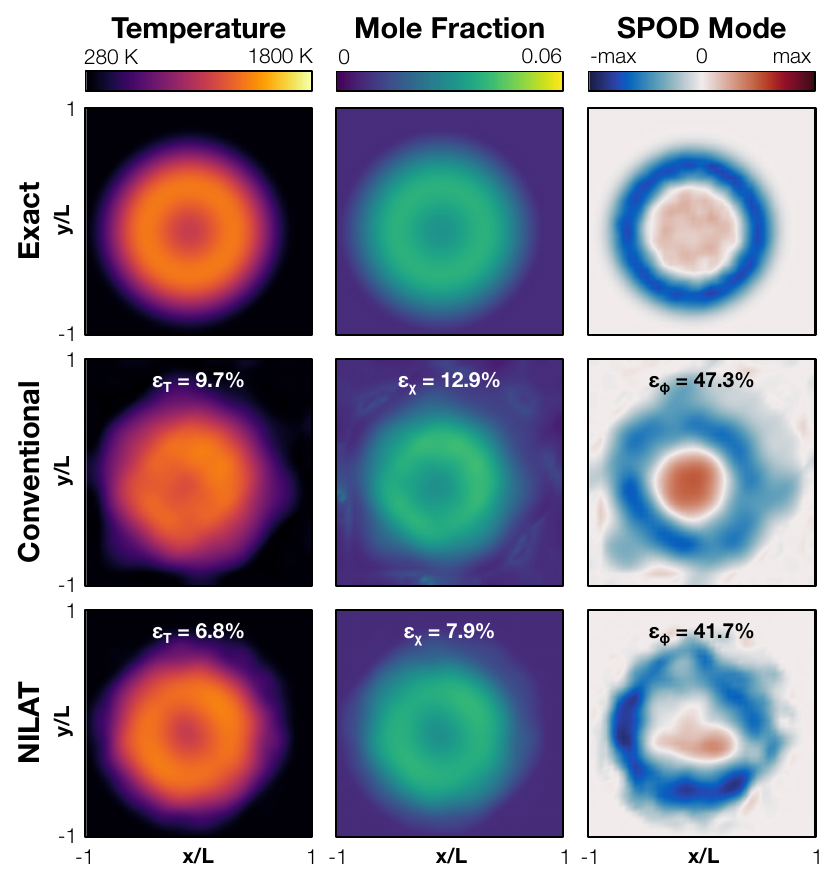}
    \caption{Reconstruction of the synthetic phantom. Rows show results from the ground truth (top), conventional method (middle), and NILAT (bottom). Columns correspond to temperature (left), \ce{H2O} mole fraction (center), and the leading SPOD mode (right).} 
    \label{fig:synthetic reconstructions}
\end{figure}

Figure~\ref{fig:synthetic reconstructions} compares the mean temperature and \ce{H2O} mole fraction fields from the ground truth phantom, the conventional reconstructions, and the NILAT reconstructions. Both methods capture the overall toroidal structure of the phantom, with peak amplitudes closely matching the true fields. However, the conventional reconstructions fail to fully resolve the cool, low-\ce{H2O} pseudo-recirculation zone at the center and exhibit noticeable artifacts near the edges of the region of interest.\par

\begin{figure}[ht]
    \centering
    \includegraphics[width=5in]{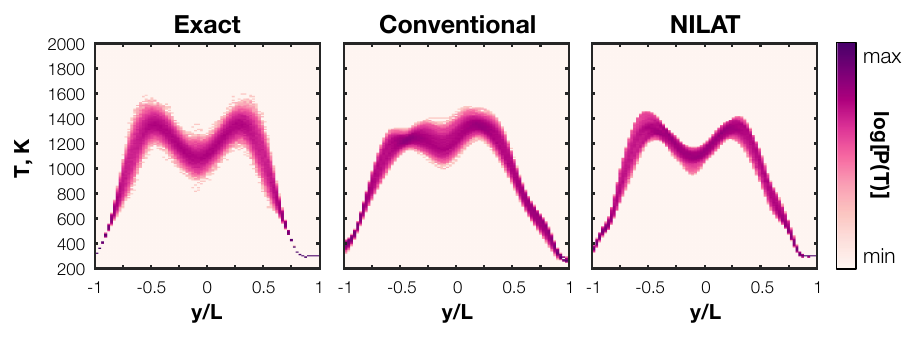}
    \caption{Normalized temperature PDFs along the vertical cut at $x = -3.15$~mm. Ground truth (left), conventional reconstruction (center), and NILAT estimate (right). NILAT more closely captures the mean profile and fluctuation structure, while the conventional result shows artifacts near the core and periphery of the phantom.}
    \label{fig:synthetic cuts}
\end{figure}

Differences between the conventional and NILAT algorithms are more pronounced when comparing probability density functions (PDFs) of the reconstructed fields. Figure~\ref{fig:synthetic cuts} presents normalized temperature PDFs extracted along a vertical cut at $x = -3.15$~mm for each method. Each 2D plot shows the normalized PDF of temperature as a function of vertical position, $y/L$. The NILAT reconstructions closely reproduce the true mean profile and capture the overall fluctuation structure, although the temperature variance is slightly underpredicted for $y > 0$, likely due to the limited spatial resolution associated with the sparse beam array.\footnote{While this study considers a fixed number of beams, prior work has demonstrated that reconstruction error generally scales with $1/m$, where $m$ is the number of laser beams \cite{Twynstra2012, Mccormick2013, McCann2022}. We verified that NILAT follows a similar trend through supplemental testing.} In contrast, the conventional reconstructions fail to capture the correct profile shape and exhibit non-physical oscillations, particularly near the center and outer edges of the phantom. These discrepancies are further illustrated in the time-resolved reconstructions provided in the supplementary material.\par

The ability of NILAT to resolve temporal dynamics is further illustrated through SPOD analysis. The first SPOD mode at the dominant frequency of 9~Hz, which captures nearly all of the coherent energy in this phantom, is shown in the top-right corner of Fig.~\ref{fig:synthetic reconstructions}. This mode features oscillations along the outer edge of the phantom, coupled with weaker, inversely correlated fluctuations near the center. NILAT recovers the mode's spatial structure well, even with limited 32-beam data, and accurately captures the mode's magnitude over time. The conventional algorithm reconstructs a qualitatively similar mode, but with noticeable distortions: the outer ring appears enlarged, an asymmetry emerges in the lower-left corner of the domain, the central structure is compressed and over-amplified, and the gradients are unrealistically sharp, likely due to reconstruction artifacts. Overall, NILAT appears well suited for reconstructing both steady-state fields and transient, coherent dynamics, offering advantages over conventional LAT algorithms in both fidelity and interpretability. Differences between the methods become more pronounced when applied to experimental data.\par

\subsection{Experimental Results}
\label{sec:results:experiment}
\begin{figure}[ht!]
    \centering
    \subcaptionbox{\label{fig:experimental reconstructions:temperature}}{
        \includegraphics[height=4.75cm]{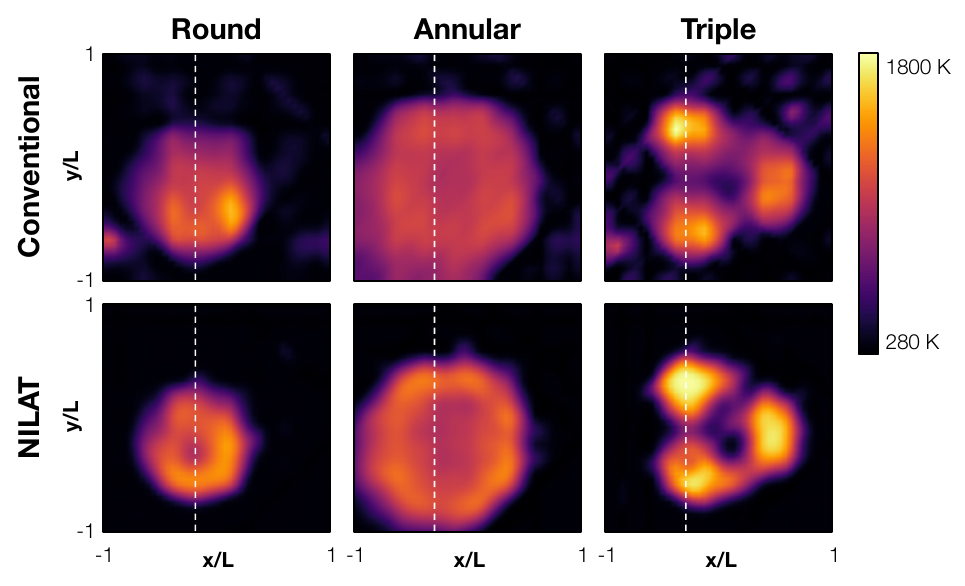}}\;
    \subcaptionbox{\label{fig:experimental reconstructions:mole fraction}}{
        \includegraphics[height=4.75cm]{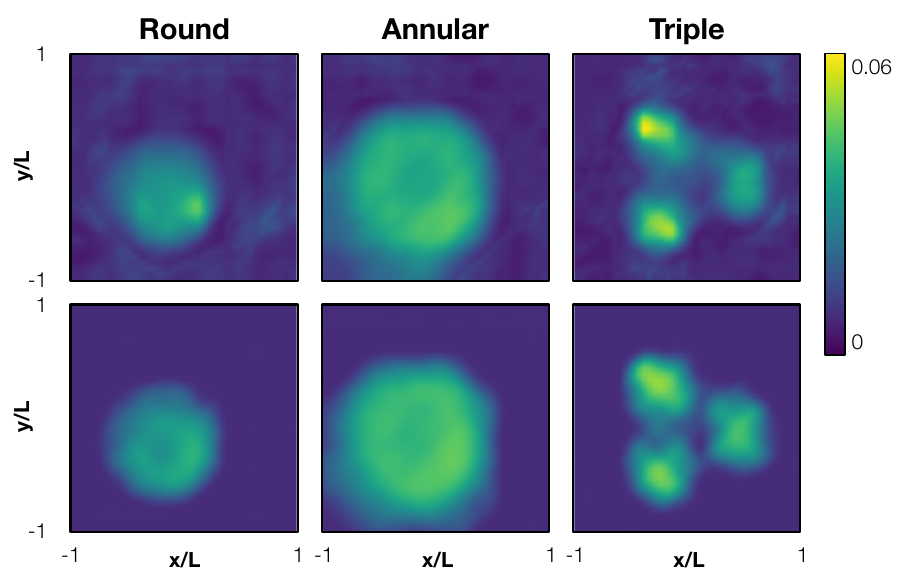}}
    \vspace*{-.66em}
    \caption{Mean reconstructions for three experimental burner configurations. (a) Temperature and (b) \ce{H2O} mole fraction fields. Dashed lines indicate the location of vertical profile cuts. LAT measurements were performed at 3~mm above the annular and triple burners and 7~mm above the round burner.}
    \label{fig:experimental reconstructions}
\end{figure}

We further demonstrate the applicability and advantages of NILAT through experimental measurements of the reacting flows described in Sec.~\ref{sec:cases:experiment}. Figure~\ref{fig:experimental reconstructions} presents the mean temperature and \ce{H2O} mole fraction fields for all three burners. The top row show results from the conventional LAT algorithm, while the bottom rows display NILAT reconstructions. Both methods recover the general structure of the combustion products, including a ring of hot water vapor above each burner cap that encloses a slightly cooler core with lower water vapor concentration. As expected, these cooler central zones become more pronounced with increasing burner cap size. Compared to the conventional approach, NILAT provides a clearer picture of these features, more accurately capturing the expected correlation between $T$ and $\chi_{\ce{H2O}}$, and producing reconstructions with sharper plume boundaries. In contrast, conventional LAT reconstructions tend to overestimate the spatial extent of the hot products, yielding smoother, more diffuse fields that obscure finer details of the flow\slash combustion processes.\par

\begin{figure}[ht!]
    \centering
    \includegraphics[width=4in]{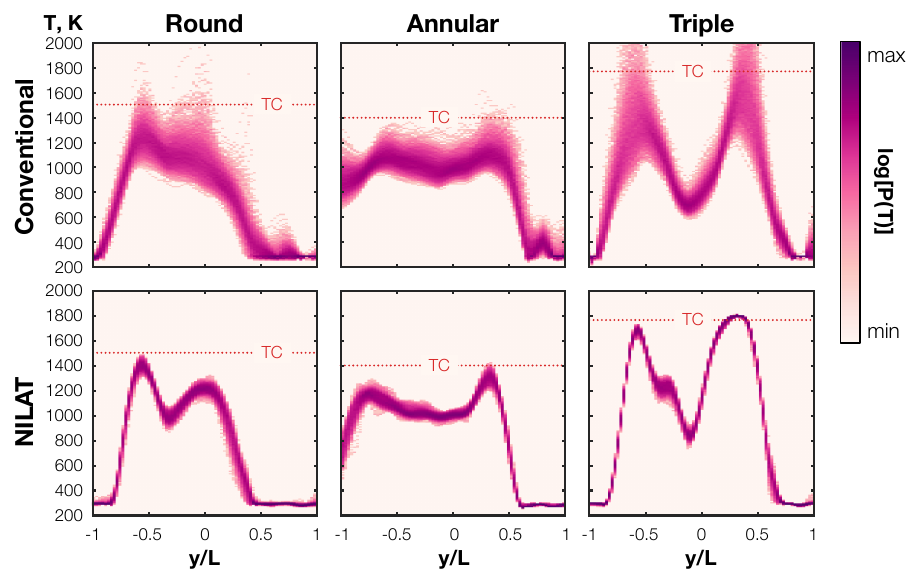}
    \caption{Reconstructed temperature PDFs from experimental data. NILAT reconstructions exhibit sharper plume structures and realistic unsteady features. In contrast, the conventional method yields spatially diffuse and erratic profiles.}
    \label{fig:experimental cuts}
\end{figure}

Normalized PDFs of temperature for the experimental reconstructions are shown in Fig.~\ref{fig:experimental cuts}, plotted along the $y/L$ axis for both the conventional algebraic method and NILAT. These PDFs highlight key differences in the reconstructed temperature distributions above each burner. The conventional reconstructions exhibit large, non-physical variances, particularly in regions that should be relatively uniform, while NILAT provides smoother, more coherent distributions that are consistent with expected flow behavior. Although direct reconstruction error cannot be assessed due to the absence of synchronous reference measurements, peak asynchronous thermocouple readings align more closely with NILAT estimates than with those from the conventional method (dashed lines in the plots). The superiority of the NILAT reconstructions is further demonstrated in the time-resolved videos provided in the supplementary material. In these videos, the conventional method exhibits non-physical temperature striations along the beam paths, whereas NILAT reconstructions are free of such artifacts. In addition to improved quantitative behavior, NILAT captures important spatial features more reliably, such as the central cold zone in the round and annular burners, and the symmetry between sub-burners in the triple configuration, further suggesting NILAT's enhanced spatial resolution.\par

\begin{figure}[ht]
    \centering
    \includegraphics[width=6in]{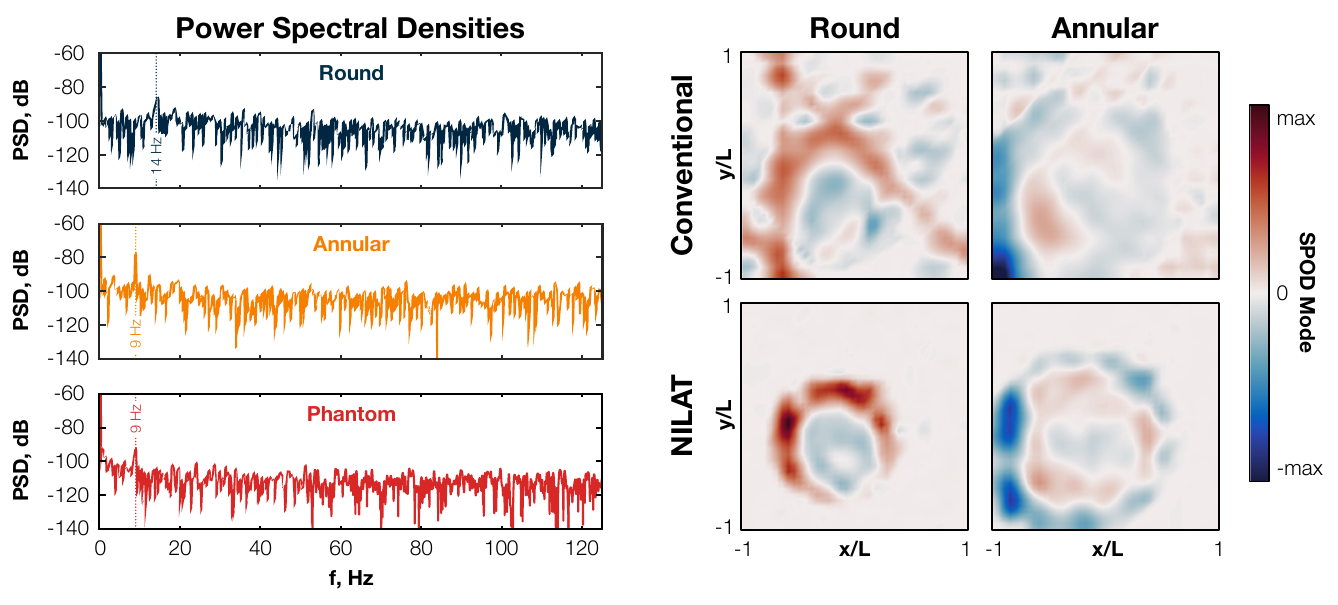}
    \caption{Spectral content and SPOD modes for the round and annular burners. PSDs (left) reveal dominant tonal modes. The corresponding SPOD modes (right) from NILAT reconstructions exhibit coherent peripheral oscillations that are anti-correlated with central fluctuations, consistent with buoyancy-driven flame flickering. In contrast, SPOD modes obtained from the conventional reconstructions display incoherent features and significant artifacts.}
    \label{fig:experimental modes}
\end{figure}

Time-resolved measurements of unsteady flames provide valuable insights into the coupling between flow and combustion processes. Power spectral density (PSD) plots in Fig.~\ref{fig:experimental modes} reveal dominant tonal frequencies of 14~Hz for the round burner, 9~Hz for the annular burner, and 9~Hz for the phantom. The phantom's tone was intentionally introduced to reflect realistic experimental dynamics, whereas the triple burner exhibited broadband fluctuations without a dominant frequency.\par

Spectral analysis of the reconstructed flow fields from the round and annular burners underscores NILAT's ability to capture coherent flame dynamics. The SPOD modes extracted at the dominant frequencies show oscillations concentrated along the plume periphery, with fluctuations at the outer edge negatively correlated with those near the center, similar to our phantom. This spatial structure is characteristic of flame flickering, a buoyancy-driven instability commonly observed in low-speed, non-premixed flames. In contrast, SPOD modes derived from conventional reconstructions appear more diffuse and lack coherent spatial organization, limiting their interpretability in this context. Flame flickering arises from buoyancy-induced vortices that form near the base of a flame due to Kelvin--Helmholtz instabilities in the shear layer \cite{Sato2000}. These vortices entrain ambient air into the reaction zone at the flame edge, locally enhancing the reaction rate and driving outward propagation of the reaction front. This cyclical entrainment and localized enhancement give rise to the periodic expansion and contraction of the flame. NILAT captures this progression, resolving both the peripheral oscillations and the corresponding central fluctuations, which are essential for interpreting the underlying flow--combustion coupling in such configurations.\par

\subsection{Computational Cost}
\label{sec:results:cost}
Neural-implicit LAT is an unsupervised learning algorithm. Unlike supervised methods that prioritize fast inference from previously trained models, NILAT requires training a new network for each dataset, adding computational costs to the reconstruction phase. This approach favors accuracy and generalizability over speed.\par

For each dataset considered here, NILAT requires approximately 3.5~hours per 2500-frame time series on an NVIDIA RTX 3090 GPU. In comparison, the conventional algebraic approach takes about 1.5~hours when parallelized over eight CPU cores (Intel Xeon W-2245). However, NILAT achieves significantly greater data compression, reducing field storage from 32~MB to 3~MB in single precision, due to its compact neural representation (approximately 315,000 trainable parameters versus over eight million in the discrete grid-based form). Moreover, NILAT's efficiency advantage becomes more pronounced in multi-transition setups. Because NILAT directly estimates temperature and mole fraction fields, it avoids separately reconstructing individual absorption coefficients and fitting spectroscopic models post hoc. This leads to sub-linear scaling in computation and storage with the number of transitions, in contrast to the linear growth observed in conventional LAT approaches (see \ref{app:LAT}).\par

\section{Conclusions}
\label{sec:conclusions}
This paper introduces \emph{NILAT}: a neural-implicit reconstruction algorithm for laser absorption tomography that estimates $\text{2D} + t$ distributions of temperature and targeted partial pressures from absorbance data. By embedding line parameters and TIPS functions into a nonlinear measurement operator, NILAT performs a direct reconstruction of the physical quantities of interest ($T$, $\chi$, $p$, etc.) rather than absorption coefficient fields. The space--time formulation supports both explicit and implicit regularization of temporal dynamics and facilitates comprehensive data assimilation. Additionally, the neural framework provides significant data compression, enabling scalability to higher spatial resolutions, longer time horizons, larger beam arrays, and multi-transition absorption setups.\par

The performance of NILAT was validated through a phantom study, where it successfully captured large-scale features of the phantom and its dynamics using a sparse imaging array. The algorithm accurately reconstructed both the toroidal temperature and water vapor mole fraction structures and the dominant temperature SPOD mode, serving as an example of high-fidelity tomographic imaging. NILAT's robustness to hysteresis effects ensures compatibility with classical parameter selection techniques like L-curve analysis. Conversely, gradient auto-weighting proved unsuitable for LAT, as the inconsistency between data and regularization loss terms led to overly smoothed solutions, a limitation not observed in PINNs, where the technique originated.\par

Experimental reconstructions using three burners further showcased NILAT's advantages. The algorithm faithfully recovered large-scale flow structures, significantly reduced artifacts, and achieved quantitative agreement with thermocouple measurements. It also effectively captured dominant flame dynamics, such as flickering, yielding SPOD modes consistent with our expectations for non-premixed flames. These findings demonstrate NILAT's potential to advance LAT applications. Future research will extend NILAT to multi-species imaging and explore its application to absorption-based velocimetry scenarios.\par

\appendix
\renewcommand{\thesection}{Appendix \Alph{section}}

\section{Discrete Laser Absorption Tomography}
\label{app:LAT}
The conventional approach to LAT begins with a vector of absorbance data in $\mathds{R}^m$,
\begin{equation}
    \mathbf{a}_k = \left(A_{k,1}, A_{k,2}, \dots, A_{k,m}\right).
    \label{equ:absorbance vector}
\end{equation}
Instead of using a coordinate-based neural network, the field variables are represented with a finite basis having $n$ functions, $\{\varphi_1, \varphi_2, \dots, \varphi_n\}$. In this work, we employ a 2D pixel basis, where the basis function $\varphi_j$ is unity inside the $j$th pixel and zero outside. An arbitrary field variable, $g$, is then approximated as:
\begin{equation}
    g\mathopen{}\left(\mathbf{x}\right) = \sum_{j=1}^n g_j \,\varphi_j\mathopen{}\left(\mathbf{x}\right),
    \label{equ:basis}
\end{equation}
where $g_j$ is the coefficient for the $j$th basis function, and $g$  represent variables such as $\chi$, $T$, or $K_k$. Field variables are thus represented by vectors of $n$ coefficients,
\begin{subequations}
    \begin{align}
        \boldsymbol\upchi &= \left(\chi_1, \chi_2, \dots, \chi_n\right) \\
        \mathbf{T} &= \left(T_1, T_2, \dots, T_n\right) \\
        \mathbf{k}_k &= \left(K_{k,1}, K_{k,2}, \dots, K_{k,n}\right),
    \end{align}
\end{subequations}
where elements of $\mathbf{k}_k$ are computed using corresponding values in $\boldsymbol\upchi$ and $\mathbf{T}$ via Eq.~\eqref{equ:int:absorption} for each $k \in \mathcal{K}$.\par

The integrated absorbance model for the $i$th beam can be approximated using the finite basis introduced above,
\begin{equation}
    A_{k,i} = \int_0^{L_i} K_k\mathopen{}\mathopen{}\left[ \mathbf{r}_i\mathopen{}\left(s\right) \right] \mathrm{d}s
    \approx \int_0^{L_i} \sum_{j=1}^{n} K_{k,j} \,\varphi_j\mathopen{}\mathopen{}\left[ \mathbf{r}_i\mathopen{}\left(s\right) \right] \mathrm{d}s
    = \sum_{j=1}^{n} K_{k,j} \underbrace{\int_0^{L_i} \varphi_j\mathopen{}\mathopen{}\left[ \mathbf{r}_i\mathopen{}\left(s\right) \right] \mathrm{d}s}_{W_{i,j} \equiv \frac{\partial A_{k,i}}{\partial K_{k,j}}},
\end{equation}
where $W_{i,j}$ is the sensitivity of the $i$th absorbance measurement to the spectral absorption coefficient in the $j$th pixel. For a pixel basis, $W_{i,j}$ is simply the chord length of the $i$th beam within the $j$th pixel; these values are assembled row-wise over beams and column-wise over pixels to form the weight matrix, $\mathbf{W}$. Given an $m \times 1$ data vector, $\mathbf{a}_k$, spectrally integrated LAT for a single transition (a.k.a. monochromatic LAT), is a linear inverse problem,
\begin{equation}
    \label{equ:matrix system}
    \mathbf{a}_k= \mathbf{Wk}_k,
\end{equation}
where $\mathbf{a}_k$ is measured and $\mathbf{k}_k$ must be inferred. This equation admits an infinite set of solutions when the column rank of $\mathbf{W}$ is less than $m$, which is guaranteed when $m < n$, as is almost always the case in LAT.\par

\subsection{Linear Reconstruction with Spectroscopic Post-Processing}
\label{app:LAT:linear}
In the linear approach to LAT, Eq.~\eqref{equ:matrix system} is inverted for each measured wavenumber or transition. The resulting values of $\mathbf{k}_k$ for $k \in \mathcal{K}$ are used to estimate the state variables at each basis function. While numerous regularization techniques exist, we focus on one of the most common methods: second-order Tikhonov regularization. This approach involves the minimization
\begin{equation}
    \label{equ:Tikhonov}
    \hat{\mathbf{k}}_k = \arg \,\underset{\mathbf{k}_k}{\min} \left\lVert\mathbf{a}_k - \mathbf{Wk}_k \right\rVert_2^2 + \gamma^2 \left\lVert \mathbf{Lk}_k \right\rVert_2^2,
\end{equation}
where $\mathbf{L}$ is the discrete Laplacian. This functional promotes smooth solutions with small second derivatives. Tikhonov regularization is computationally efficient and generally produces reasonable results, but the formulation in Eq.~\eqref{equ:Tikhonov} lacks a direct connection to the spatial derivatives of $\chi$ and $T$.\par

For spectrally integrated data, local Boltzmann plots are used to determine $\chi$ and $T$. These plots incorporate reconstructed absorption coefficient values and line parameters,
\begin{subequations}
    \label{equ:Boltzmann}
    \begin{align}
        \label{equ:Boltzmann:y}
        y_\mathrm{B} &= \underbrace{\log\mathopen{}\left[\frac{K_k}{S_{\mathrm{ref},k}} \exp\mathopen{}\left(\frac{c_2 E_k^{\prime\prime}}{T_\mathrm{ref}}\right) \right]}_\text{measured and plotted}
        = \underbrace{\frac{-c_2 E_k^{\prime\prime}}{T} + \log\mathopen{}\left[\frac{\chi p}{k_\mathrm{B} T} \frac{Q\mathopen{}\left(T_\mathrm{ref}\right)}{Q\mathopen{}\left(T\right)}\right]}_\text{unknown $
        \{\chi, T, p\}$} \\
        \label{equ:Boltzmann:x}
        x_\mathrm{B} &= c_2 E_k^{\prime\prime},
    \end{align}
\end{subequations}
where each transition at each basis function (pixel) provides one $(x_\mathrm{B}, y_\mathrm{B})$ point.\footnote{The simplified expression for $y_\mathrm{B}$ is obtained by substituting Eqs.~\eqref{equ:int:absorption} and \eqref{equ:line intensity} into Eq.~\eqref{equ:Boltzmann:y}, assuming that $c_2 \nu_k \ll 1$, which is reasonable at the wavenumbers and temperatures considered in this work.} Using this definition,
\begin{subequations}
    \label{equ:Boltzmann analysis}
    \begin{align}
        T &= -\left(\frac{\mathrm{d}y_\mathrm{B}}{\mathrm{d}x_\mathrm{B}}\right)^{-1}
        \intertext{and}
        \chi &= \frac{k_\mathrm{B} T}{p} \frac{Q\mathopen{}\left(T\right)}{Q\mathopen{}\left(T_\mathrm{ref}\right)} \exp\mathopen{}\left(\frac{c_2 E_k^{\prime\prime}}{T}\right).
    \end{align}
\end{subequations}
These expressions are evaluated at each pixel, and accuracy improves with increasing spectral information. This approach reduces to ratiometric thermometry when only two transitions are available. In the spectrally-resolved case, the local thermochemical state is determined through regression, as described in \cite{Grauer2019}.\par

\subsection{Spectrally Integrated Nonlinear Reconstruction}
\label{app:LAT:nonlinear}
The nonlinear LAT reconstruction problem with second-order Tikhonov regularization for the mole fraction and temperature fields corresponds to the following minimization:
\begin{equation}
    \label{equ:functional}
    \left(\hat{\boldsymbol\upchi}, \hat{\mathbf{T}}\right) = \mathrm{arg}\,\underset{\left(\boldsymbol\upchi, \mathbf{T}\right)}{\mathrm{min}} \left.
    \sum_{k \in \mathcal{K}} \left\lVert \mathbf{a}_k - \mathbf{W} \,\mathbf{k}_k\mathopen{}\left(\boldsymbol\upchi, \mathbf{T}\right) \right\rVert_2^2 +
    \gamma_\upchi^2 \left\lVert \mathbf{L}\boldsymbol\upchi \right\rVert_2^2 +
    \gamma_\mathrm{T}^2 \left\lVert \mathbf{LT} \right\rVert_2^2 \right.,
\end{equation}
which can be solved using a variety of optimization techniques. Note that we have not introduced any time dependencies in our presentation of the conventional LAT problem. While it is possible to perform space--time reconstructions using a discrete formulation, the dimensions of $\boldsymbol\upchi$ and $\mathbf{T}$  increase linearly with the number of time steps, resulting in very large matrix systems. In contrast, $\mathsf{N}$ offers a highly compressed representation of $(\chi, T)$, making it well-suited for long datasets.\par

\section{Network Architecture}
\label{app:network}
In NILAT, coordinate neural networks are used to represent the gas state as a function of space and time. The network maps input coordinates, $\mathbf{z}^0 = (\mathbf{x}, t)$, to outputs, $\mathbf{z}^{n_\mathrm{l}+1} = (\chi, T)$, through a series of $n_\mathrm{l}$ hidden layers,
\begin{subequations}
    \label{equ:method:architecture}
    \begin{align}
        \mathbf{z}^{n_\mathrm{l} + 1} &= \mathsf{N}\mathopen{}\left(\mathbf{z}^0\right) = \mathrm{sigmoid}\mathopen{} \left\{\mathbf{W}^{n_\mathrm{l}+1}\mathopen{}\left[\mathsf{L}^{n_\mathrm{l}} \circ \mathsf{L}^{n_\mathrm{l}-1} \circ \cdots \circ \mathsf{L}^2 \circ \mathsf{F}\mathopen{}\left(\mathbf{z}^0\right)\right]+\mathbf{b}^{n_\mathrm{l}+1}\right\},
        \intertext{where the standard layers, $\mathsf{L}^l$, have the following structure:}
        \mathbf{z}^l &= \mathsf{L}^l\mathopen{}\left(\mathbf{z}^{l-1}\right) = \mathrm{swish}\mathopen{}\left(\mathbf{W}^l\mathbf{z}^{l-1} + \mathbf{b}^l\right) \quad\text{for}\quad l \in \{2, 3, \dots, n_\mathrm{l}\}.
    \end{align}
\end{subequations}
Here, $\mathbf{W}^l$ and $\mathbf{b}^l$ are the weight matrix and bias vector for the $l$th layer and \begin{subequations}
    \begin{align}
        \mathrm{sigmoid}\mathopen{} \left(z_i\right) &= \frac{1}{1 + \exp\mathopen{} \left(-z_i\right)} \\
        \mathrm{swish}\mathopen{} \left(z_i\right) &= z_i \,\mathrm{sigmoid}\mathopen{} \left(z_i\right)
    \end{align}
\end{subequations}
are activation functions, which are applied element-wise to vector or matrix inputs. The swish activation is smooth and avoids saturation, making it well-suited for hidden layers in a coordinate neural network. The sigmoid activation on the final layer ensures non-negative outputs, which are then linearly transformed to lie within prescribed physical ranges. All weights and biases are collected in the trainable parameter vector, $\boldsymbol\uptheta$, which is updated by minimizing $\mathscr{J}_\mathrm{total}$.\par

To enhance spectral resolution, the first layer, $\mathsf{L}^1$, is replaced with a Fourier encoding layer \cite{Tancik2020},
\begin{equation}
    \label{equ:method:FE}
    \mathbf{z}^1 = \mathsf{F}\mathopen{}\left(\mathbf{z}^0\right) = \mathopen{}\left[\sin\mathopen{}\left(2\pi \mathbf{f}_1 \cdot \mathbf{z}^0\right), \,\cos\mathopen{}\left(2\pi \mathbf{f}_1 \cdot \mathbf{z}^0\right), \dots, \,\sin\mathopen{}\left(2\pi \mathbf{f}_{n_\mathrm{f}} \cdot \mathbf{z}^0\right), \,\cos\mathopen{}\left(2\pi \mathbf{f}_{n_\mathrm{f}} \cdot \mathbf{z}^0\right)\right],
\end{equation}
where $\mathbf{f}_i$ are randomly sampled frequencies (see \ref{app:Fourier}). This mitigates the low-frequency spectral bias of gradient-descent-based training \cite{Wang2021}.\par

Neural reconstructions were implemented in PyTorch~2.0.1 using a 1024-feature Fourier encoding and five hidden layers with 250 nodes each. Weights were initialized from a standard normal distribution and biases were set to zero. Outputs from the sigmoid activation were mapped to physical bounds: $0.0075 \leq \chi < 0.053$ and $280~\text{K} \leq T < 1800~\text{K}$. For other applications, these limits can be adjusted based on known thermodynamic constraints, e.g., an adiabatic flame temperature. Inputs were normalized by the spatial and temporal extent of the dataset; outputs were range-normalized and dimensionalized before evaluating the data loss in Eq.~\eqref{equ:data loss}. The regularization loss in Eq.~\eqref{equ:regularization loss} was computed in non-dimensional form for numerical stability.\par

Reconstructions were trained using the Adam optimizer over 80 epochs with a learning rate of  $10^{-3}$, followed by four refinement epochs at a rate of $10^{-4}$. All reconstructions spanned the full octagonal sensing region, with ambient conditions weakly enforced on  $\partial\mathcal{A}$ via the boundary loss in Eq.~\eqref{equ:boundary}. The boundary was defined dynamically as the largest ellipse enclosed by beams that did not intersect hot gases. Each training batch included all beams at five time instances, evaluated for all transitions in $\mathscr{J}_\mathrm{data}$, 10,000 interior points for $\mathscr{J}_\mathrm{penalty}$, and 10,000 ambient points for $\mathscr{J}_\mathrm{bound}$. Absorbances were computed using 2000 random integration points per beam path, yielding relative errors below 2\%.\par

\section{Fourier Encoding Formulation}
\label{app:Fourier}
Fourier encodings are essential in NILAT for reconstructing unsteady, spatially complex flow fields. This appendix illustrates three key aspects of their role. First, the encodings are necessary to represent fields with complex spatio-temporal structures. Second, explicit regularization becomes essential once an encoding has been introduced. Third, the accuracy of reconstructions depends on the frequency distribution used to generate the encoding features, particularly the temporal component for tonal flows. These findings are supported by reconstruction tests using the synthetic phantom from Sec.~\ref{sec:cases:phantom}, which features broadband fluctuations and a dominant tone at 9~Hz.\par

Each Fourier encoding is constructed by drawing frequency vectors, $\mathbf{f}$, corresponding to the space--time input, $\mathbf{z}^0 = (x, y, t)$. The spatial components, $f_1$ and $f_2$, are drawn from a zero-mean Gaussian with a standard deviation of $\sigma_\mathbf{x} = 0.5$~cm$^{-1}$. Temporal frequencies, $f_3$, are drawn from a probability density function, $P(f_3)$, which may be a unimodal or multimodal Gaussian mixture,
\begin{equation}
    P\mathopen{} \left(f_3\right) = \sum_{i=1}^N \frac{w_i}{\sqrt{2\pi \sigma_i^2}} \exp\mathopen{}\left[-\frac{\left(f_3 - \mu_i\right)^2}{2\sigma_i^2}\right], \quad \sum_{i=1}^N w_i = 1.
\end{equation}
We consider three unimodal distributions with $\mu = 0$~Hz and $\sigma = 5, 10, \text{and} \ 15$~Hz. We also consider a trimodal distribution with a central peak at 0~Hz ($\sigma = 0.2$~Hz, $w = 0.5$) and two side peaks centered at $\pm f_\mathrm{flow}$ (i.e., 9~Hz for the phantom), each with $\sigma_2 = \sigma_3 = 1$~Hz and $w_2 = w_3 = 0.25$. This formulation is inspired by the approach of Jin et al. \cite{Jin2024} and allows the encoding to reflect a priori knowledge of the system's frequency content. The dominant flow frequency, $f_\mathrm{flow}$, may be determined in practice from the PSD of the measured absorbance data (see Fig.\ref{fig:experimental modes}).\par

\begin{figure}[ht!]
    \centering
    \includegraphics[width=6.25in]{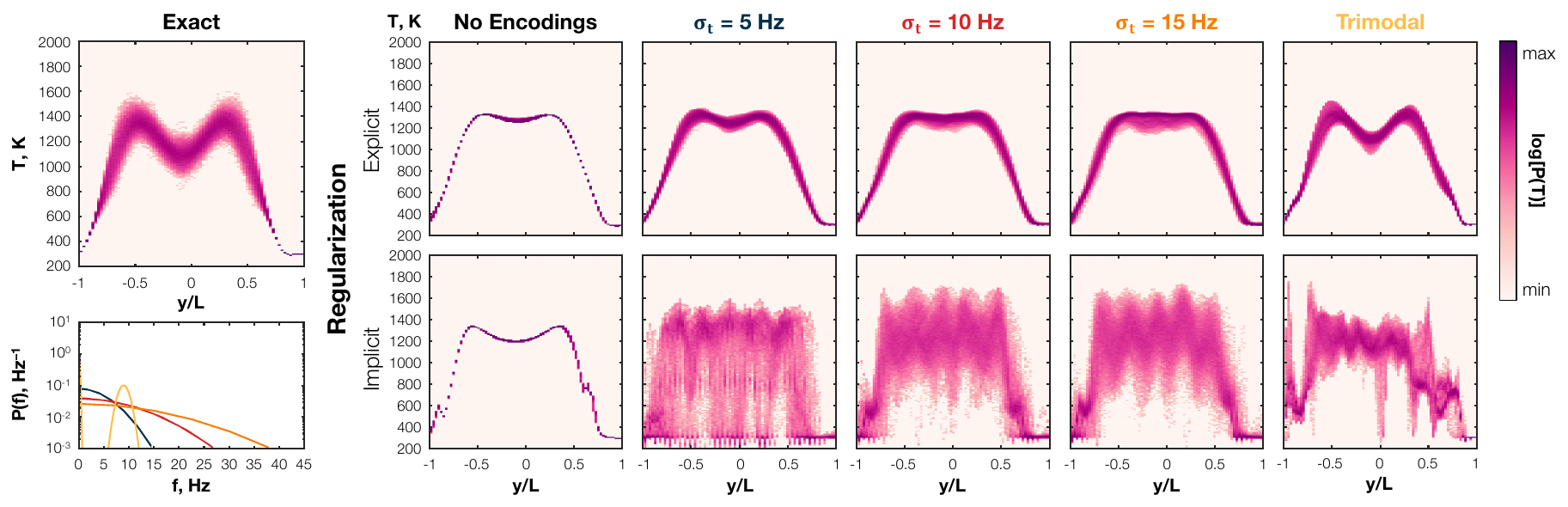}
    \caption{Impact of Fourier encodings and regularization. (Left panel) Distributions used to draw temporal encoding frequencies, overlaid with the ground-truth temperature PDF along a vertical cut. (Right panel) Reconstructed temperature PDFs using different encodings (columns) and either explicit (top row) or implicit (bottom row) regularization.}
    \label{fig:Fourier}
\end{figure}

Figure~\ref{fig:Fourier} visualizes the effects of Fourier encodings and regularization. The left side shows the temporal frequency PDFs used to draw $f_3$, plotted below the ground-truth temperature PDF along the vertical cut at $x = -3.15$~mm. The right side presents reconstructed temperature PDFs using various encoding strategies and regularization settings. Columns correspond to different encodings, from no encoding (leftmost), to unimodal Gaussians, to the trimodal distribution (rightmost). Rows indicate the use of an explicit penalty term (top) versus implicit regularization only (bottom).\par

This figure illustrates the three central findings described above. First, networks without Fourier encodings fail to recover fluctuations, as can be seen in the first column of estimated temperature PDFs. The standard MLPs produce nearly static reconstructions and misestimate even the mean profile due to the nonlinear spectroscopic model. The bottom-left case loosely corresponds to the approach of Li et al. \cite{Li2024}, which omits both encodings and regularization. Second, while Fourier encodings enable the network to represent unsteady fields, they must be paired with explicit regularization. Without a regularization term, encoding-enhanced networks exhibit high-frequency artifacts due to increased expressivity. Explicit penalties, such as Tikhonov regularization, suppress these spurious modes and yield physically plausible results. Third, the frequency distribution used in the encoding significantly affects reconstruction accuracy. Increasing the width of unimodal distributions does not consistently improve performance and may destabilize training. Conversely, tailoring the encoding distribution to reflect dominant flow frequencies (e.g., identified from the measurement PSDs) yields notable improvements. This is evident in the upper-right plot of Fig.~\ref{fig:Fourier}, where the trimodal encoding leads to the most accurate recovery of both mean and fluctuating temperature fields.\par

\section*{Novelty and Significance Statement}
Industrial environments, such as gas turbine test beds, present significant diagnostic challenges due to harsh operating conditions and limited optical access. In this work, we demonstrate the first long-time-horizon reconstructions of simultaneous 2D temperature and water vapor mole fraction fields in laboratory burners using neural-implicit laser absorption tomography (NILAT). We characterize NILAT's performance through a synthetic phantom study featuring a realistic mean profile, broadband fluctuations, and tonal dynamics, highlighting its robustness and reconstruction accuracy. We also validate the applicability of established regularization parameter selection methods. This sensing framework extends beyond controlled laboratory conditions and offers potential for deployment in extreme environments where direct measurements are impractical.\par

\section*{Declaration of Competing Interests}
The authors declare that they have no known competing financial interests or personal relationships that could have appeared to influence the work reported in this paper.\par

\section*{Acknowledgments}
C.L. acknowledges support from the EPSRC through Programme Grant EP/T012595/1, Platform Grant EP/P001661/1, and Impact Acceleration Account PV120. S.J.G. acknowledges support from NASA under contract 80NSCC24PB449 and from FAU Erlangen-N{\"u}rnberg. J.X. acknowledges support from the Worshipful Company of Instrument Makers through a Postgraduate Scholarship. J.P.M. acknowledges support from the DoD through an NDSEG Fellowship.\par


\end{document}